\documentclass[a4paper]{article}
\usepackage[pdftex]{graphicx}  

\pdfoutput=1

\begin{document}

\title{\bf
A fast method to estimate speciation parameters in a model of isolation with an initial period of gene flow \\
and to test alternative evolutionary scenarios.}
\author{\\Hilde M. Wilkinson-Herbots
\\Department of Statistical Science, 
University College London\\ Gower Street, London WC1E 6BT, UK \\
email: h.herbots@ucl.ac.uk}

\date{}

\maketitle

\begin{abstract}
\noindent
We consider a model of ``isolation with an initial period of migration" (IIM), where an ancestral population instantaneously split into two descendant populations which exchanged migrants symmetrically at a constant rate for a period of time but which are now completely isolated from each other.
A method of Maximum Likelihood estimation of the parameters of the model 
is implemented,
for data consisting of the number of nucleotide differences between two DNA sequences at each of a large number of independent loci, using the explicit analytical expressions for the likelihood obtained in Wilkinson-Herbots~(2012).
The method is demonstrated on a large set of DNA sequence data from two species of Drosophila,
as well as on simulated data.
The method is extremely fast, returning parameter estimates in less than 1 minute for a data set 
consisting of the numbers of differences between pairs of sequences from 10,000s of loci,
or in a small fraction of a second if all loci are trimmed to the same estimated mutation rate. 
It is also illustrated how the maximized likelihood can be used to quickly distinguish between competing models describing alternative evolutionary scenarios, either by comparing AIC scores or by means of likelihood ratio tests. The present implementation is for a simple version of the model, but various extensions are possible and are briefly discussed.
\end{abstract}
\vspace*{0.5cm}

\section*{Introduction}

In recent years, molecular genetic data have been used extensively to learn about 
the evolutionary processes that gave rise to the observed genetic variation. 
One important example is the use of 
genetic data to try to infer whether or not gene flow occurred 
between closely related species during or after speciation (see for 
example reviews by Nadachowska~2010, Pinho and Hey~2010, 
Smadja and Butlin~2011, Bird et al.~2012 and a large number of references 
in these papers).
Such applications typically use computer programs such as  
{\em MDIV} (Nielsen and Wakeley~2001), {\em IM} (Hey and Nielsen~2004, 
Hey~2005), {\em IMa} (Hey and Nielsen~2007), 
{\em MIMAR} (Becquet and Przeworski~2007) or 
{\em IMa2} (Hey~2010), based on the ``isolation with migration" (IM)
model, which assumes that a panmictic ancestral population instantaneously split 
into two or more descendant populations some time in the past and that 
migration occurred between these descendant populations at a constant rate 
ever since.
Whilst such methods have been extensively and successfully applied to study 
the relationships
between different populations within species, the assumption of migration
continuing at a constant rate until the present is unrealistic when 
studying relationships between species. 
Becquet and Przeworski~(2009) and Strasburg and Rieseberg~(2010) 
investigated by means of simulations 
how robust parameter estimates (migration rates, divergence times and 
effective population sizes), obtained by methods based on the IM model,
are to violations of the IM model assumptions. 
Becquet and Przeworski~(2009) found that parameter estimates
obtained with {\em IM} and {\em MIMAR} 
are often biased when the assumptions of the IM model are violated,
and concluded that these methods are highly sensitive to the assumption of a constant
migration rate since the population split.
Theoretical results derived by Wilkinson-Herbots~(2012) also suggested that
estimated levels of gene flow obtained by applying an IM model to species 
which are now completely isolated cannot simply be interpreted as average 
levels of gene flow over time. 
Thus, whilst computer programs based on the IM model can be used to test for
departure from a complete isolation model (which assumes that an ancestral 
population instantaneously split into
two or more descendant populations which 
have been completely isolated ever since),
the actual parameter estimates obtained may be difficult to interpret if the 
IM model is not an accurate description 
of the history of the populations or species concerned.
Furthermore, in some studies the programs {\em IMa} or {\em IMa2} have also 
been used
to estimate the times when migration events occurred, and to try to 
distinguish between scenarios of speciation with gene flow and scenarios 
of introgression, 
and it has recently been demonstrated   
that such inferences about the timing of gene flow 
(obtained with programs based on the IM model)
are not valid 
(Strasburg and Rieseberg~2011, Sousa et al.~2011).
Becquet and Przeworski~(2009), Strasburg and Rieseberg~(2011) and 
Sousa et al.~(2011) all suggested that more realistic models
of population divergence and speciation are needed. 

A first step in trying to make the IM model more suitable for the study of 
speciation is to allow gene flow to occur during a limited period of time, 
followed by complete isolation of the species, and such models 
have been studied by Teshima and Tajima~(2002),
Innan and Watanabe~(2006) 
and Wilkinson-Herbots~(2012).
In the latter paper, a model of ``isolation with an initial period of 
migration" (IIM) was studied where a panmictic ancestral population gave 
rise to two or more descendant populations which exchanged
migrants symmetrically at a constant rate for a period of time, after which
they became completely isolated from each other. Explicit analytical  
expressions were derived for the probability that two DNA sequences (from 
the same descendant population or from different descendant 
populations) differ at $k$ nucleotide sites, assuming the infinite sites
model of neutral mutation. It was suggested that these results may be 
useful for Maximum Likelihood estimation of the parameters of the model,  
if one pair of DNA sequences is available at each of a large number
of independent loci, and that such an ML method should be very fast as  
it is based on an explicit expression for the likelihood rather than on
computation by means of numerical approximations or MCMC simulation.
However the proposed ML estimation method had 
not yet been implemented and its usefulness was 
still to be demonstrated.
In this paper we present an implementation of the ML estimation method 
for the parameters of the ``isolation with an initial period of migration" 
model proposed 
in Wilkinson-Herbots~(2012), and we illustrate its potential by
applying it to a set of DNA sequence data from {\em Drosophila simulans}  
and {\em Drosophila melanogaster} previously analysed by Wang and Hey~(2010)
and Lohse et al.~(2011). The parameter estimates and maximized likelihood obtained for the 
IIM model are also compared to those under an IM model and an isolation model,
and it is illustrated that the method makes it possible to distinguish 
between more and less plausible evolutionary scenarios, by comparing AIC scores or by means of likelihood ratio tests. 
The implementation was done in R (R Development Core Team 2011). 
The R code for the MLE method described is included as Supplementary Material 
and can readily be pasted into an R document.
This MLE algorithm is very fast indeed: for the Drosophila  data mentioned 
above (using the number of nucleotide differences between two DNA sequences
at each of approximately 30,000 loci), we obtained estimates of the 
parameters of the IIM model using a desktop PC; the
program returned results instantly (i.e. in a small fraction of a second) if all loci had been 
trimmed to the same estimated mutation rate (as proposed by Lohse et al.
2011), or in approximately 20 seconds
if the full sequences (and hence different mutation rates at different loci) 
were used.

Whereas most earlier methods used large samples from a small number of loci,
recent advances in DNA sequencing technology and the advent of whole-genome
sequencing have led to an increased interest in methods which (like that
described in this paper) can handle
data from a small number of individuals 
at a large number of loci 
(for example, Takahata~1995; Takahata et al.~1995; Takahata and Satta~1997; 
Yang~1997, 2002; Innan and Watanabe~2006; Wilkinson-Herbots~2008; 
Burgess and Yang~2008; Wang and Hey~2010; Yang~2010; Hobolth et al.~2011; 
Li and Durbin~2011; Lohse et al.~2011; Wilkinson-Herbots~2012;
Zhu and Yang~2012; Andersen et al.~2014). 
This type of data set has two advantages.
Firstly, data from even a very large number of individuals at the same 
locus tend to contain only little information about very old divergence or 
speciation events 
because typically the individuals' ancestral lineages will have 
coalesced 
to a very small number of ancestral lineages 
by the time the event of interest is reached,
and in such contexts
a data set consisting of a small number of DNA sequences at each of a 
large number of independent loci 
may be more informative (Maddison and Knowles~2006; Wang and Hey~2010; 
Lohse et al.~2010, 2011).
Secondly, considering small numbers of sequences at large numbers of 
independent loci is mathematically much easier and computationally much 
faster than working with large numbers of sequences at the same locus. 
In particular, explicit analytical expressions for the likelihood have been 
obtained for pairs or triplets of sequences for a number of demographic 
models (for example, Takahata et al.~1995; Wilkinson-Herbots~2008; 
Lohse et al.~2011; Wilkinson-Herbots~2012; Zhu and Yang~2012), which substantially
speeds up computation and maximization of the likelihood.

\section*{New Approaches}

\begin{figure}[t]
\centering
\vspace*{-2cm}
\includegraphics[width=10.5cm, angle=0]{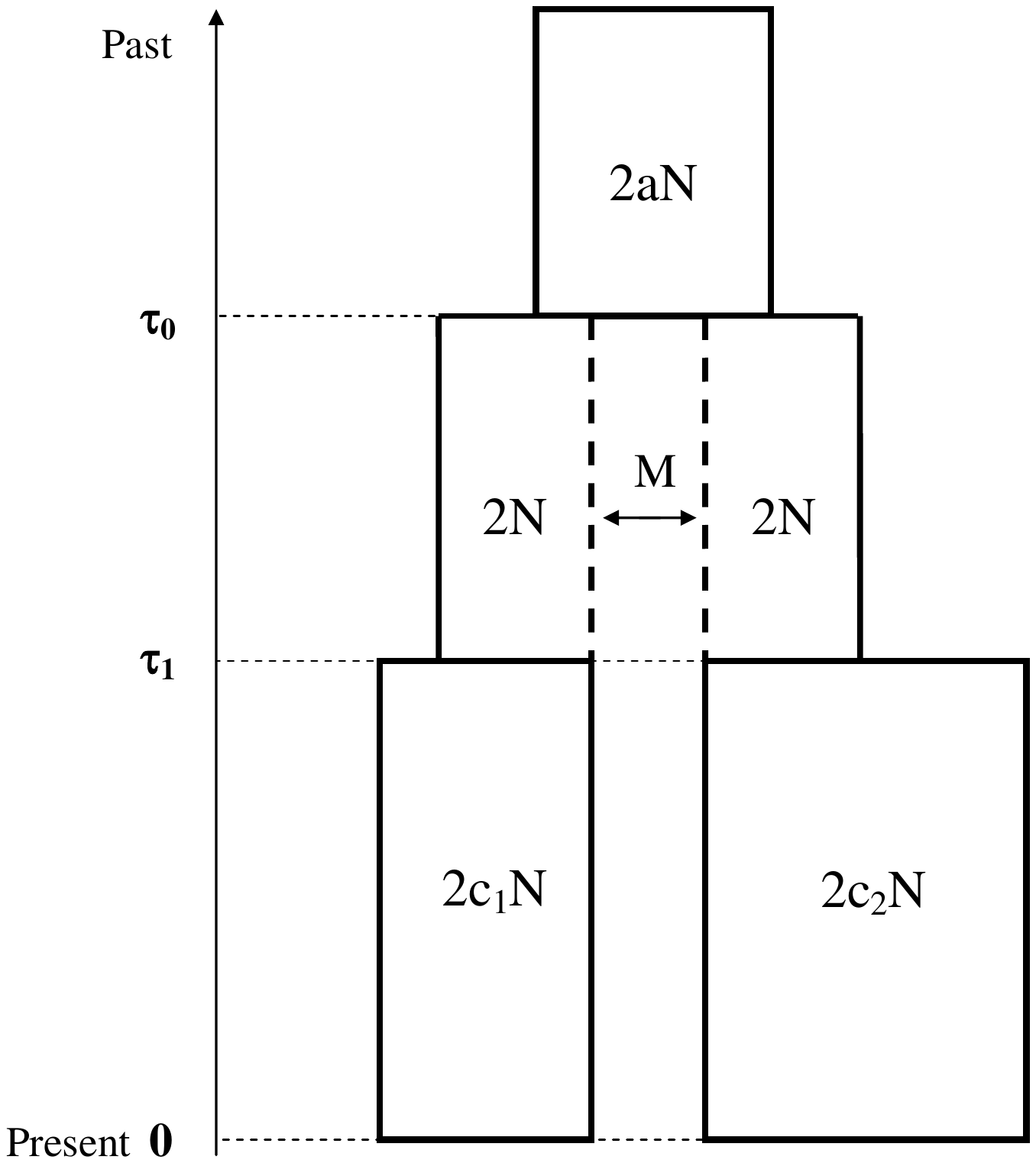}
\vspace{-5.3cm}
\caption{ The ``isolation with initial migration" (IIM) model.}
\label{fig1}
\end{figure}

We obtain Maximum Likelihood estimates of the parameters of the 
``isolation with initial migration" (IIM) model studied in 
Wilkinson-Herbots~(2012), for the case of $n=2$ descendant populations or
species. This model assumes that, time $\tau_0$ ago 
($\tau_0>0$), a panmictic ancestral population instantaneously split into  
two descendant populations which subsequently exchanged 
migrants symmetrically at a constant rate until time $\tau_1$ ago 
($0<\tau_1<\tau_0$), when they became completely isolated from each other
(see Figure~\ref{fig1}). 
Focusing on DNA sequences at a single locus that is not subject to intragenic 
recombination, the ancestral population is assumed to have been of constant 
size $2aN$ homologous 
sequences until the split occurred time $\tau_0$ ago, where $N$ is large.
Between time $\tau_0$ ago and time $\tau_1$ ago, the two descendant 
populations were of constant size $2N$ sequences each and 
exchanged migrants at a constant rate $m$, where $m$ is the 
proportion of each descendant population that was replaced by immigrants each 
generation. The current size of descendant population~$i$ is $2c_iN$  
sequences ($i=1,2$), and is
assumed to have been constant since migration ended time $\tau_1$ ago.
As is standard in coalescent theory and assuming that reproduction within 
populations follows the neutral Wright-Fisher model,
time is measured in units of $2N$ generations
(this also applies to the times $\tau_0$ and $\tau_1$);
in practical applications where the Wright-Fisher model does 
not hold, $N$ is interpreted as the effective population size. 
The ``scaled" migration and mutation rates 
are defined as $M=4Nm$ and $\theta=4N\mu$, respectively, where $\mu$ is the 
mutation rate per DNA sequence per generation at the locus concerned.
The work described in this paper assumes 
that mutations are selectively neutral and follow the infinite sites model 
(Watterson~1975). Extensions to other neutral mutation models are
feasible but have not yet been implemented.

For this IIM model, Wilkinson-Herbots~(2012) found the probability that 
two homologous DNA sequences differ at $k$ nucleotide sites:
denoting by $S_{ij}$ the number of nucleotide differences between 
two homologous sequences sampled from descendant populations~$i$ and~$j$ 
($i,j \in \{1,2\}$)
and using the subscript $\theta$ to indicate the scaled mutation rate at the 
locus concerned, we have for $k=0,1,2,\ldots$
\begin{eqnarray}
P_\theta(S_{ii}=k) & = & 
 \frac{(c_i \theta)^k}{(1 + c_i \theta)^{k+1}} \left( 1 -
e^{-\tau_1 (\frac{1}{c_i} + \theta)} 
\sum_{l=0}^{k} \frac{ \left( \frac{1}{c_i} + \theta \right)^l \tau_1^l }{l!}
\right) \nonumber \\ 
 & & \mbox{} + e^{-\frac{1}{c_i}\tau_1} \sum_{r=1}^{2} A_{0r} 
 \frac{\lambda_r \theta^k}{(\lambda_r+\theta)^{k+1}} \left(
e^{-\theta \tau_1} \sum_{l=0}^{k} \frac{(\lambda_r+\theta)^l \tau_1^l}{l!}
- e^{-\lambda_r (\tau_0 - \tau_1) - \theta \tau_0} 
\sum_{l=0}^{k} \frac{(\lambda_r + \theta)^l \tau_0^l}{l!}
 \right) \nonumber \\ 
 & & \mbox{} + \frac{e^{-\frac{1}{c_i}\tau_1-\theta \tau_0} (a \theta)^k}{
 (1 + a \theta)^{k+1}} \left( 
 \sum_{l=0}^{k} \frac{ \left( \frac{1}{a} + \theta \right)^l \tau_0^l }{l!}
\right) \sum_{r=1}^{2} A_{0r} e^{ - \lambda_r (\tau_0 - \tau_1)} 
\label{distribution0}
\end{eqnarray}
for a pair of sequences from the same descendant population,
and \\
\begin{eqnarray}
P_\theta(S_{12}=k) 
& = & \sum_{r=1}^{2} A_{1r} 
\frac{\lambda_r \theta^k}{(\lambda_r+\theta)^{k+1}} \left(
e^{-\theta \tau_1} \sum_{l=0}^{k} \frac{(\lambda_r+\theta)^l \tau_1^l}{l!}
- e^{-\lambda_r (\tau_0 - \tau_1) - \theta \tau_0} 
\sum_{l=0}^{k} \frac{(\lambda_r + \theta)^l \tau_0^l}{l!}
 \right) \nonumber \\
 & & \mbox{} + \frac{e^{-\theta \tau_0} (a \theta)^k}{
 (1 + a \theta)^{k+1}} \left( 
 \sum_{l=0}^{k} \frac{ \left( \frac{1}{a} + \theta \right)^l \tau_0^l }{l!}
\right) \sum_{r=1}^{2} A_{1r} e^{ - \lambda_r (\tau_0 - \tau_1)} 
\label{distribution1}
\end{eqnarray}
for a pair of sequences from different descendant populations,
where
$$
\begin{array}{ll}
\lambda_1 = {\displaystyle \frac{nM+n-1-\sqrt{D}}{2(n-1)}}~~~~~~~~~~~~~~~~~ &
\lambda_2 = {\displaystyle \frac{nM+n-1+\sqrt{D}}{2(n-1)}} 
\end{array}
$$
with  
$$
D = (nM+n-1)^2 - 4(n-1)M
$$
and where 
$$
\begin{array}{ll}
A_{01} = {\displaystyle \frac{\lambda_2-1}{\lambda_2-\lambda_1}}~~~~~~~~~~~~~~~~~~~~~~~~~~~~~~~ & 
A_{02} = {\displaystyle \frac{1-\lambda_1}{\lambda_2-\lambda_1}}\\ 
A_{11} = {\displaystyle \frac{\lambda_2}{\lambda_2-\lambda_1}} &
A_{12} = {\displaystyle \frac{-\lambda_1}{\lambda_2-\lambda_1}}. 
\end{array}
$$
Thus for pairwise difference data of the form 
$\{k_l; l=1,\ldots,K\}$, consisting of the number of 
nucleotide differences between one pair of DNA sequences 
at each of $K$ independent loci, the likelihood under the IIM model 
is given by 
\begin{equation}
\prod_{l=1}^{K} P_{\theta^{(l)}}(S_{i_l j_l}=k_l)
\label{likelihood}
\end{equation}
where 
$i_l$ and $j_l$ denote the locations (i.e. the population 
labels) of the two DNA sequences sampled at the $l$th locus, and
where each factor $P_{\theta^{(l)}}(S_{i_l j_l}=k_l)$ is given by 
equation~(\ref{distribution0}) or~(\ref{distribution1}) as 
appropriate, replacing $\theta$ by the scaled mutation rate
$\theta^{(l)}$ for the $l$th locus. 
Note that the above derivation of the likelihood assumes that there is no recombination within loci and free recombination between loci.
In order to jointly estimate all the 
parameters of the model, data from between-population sequence comparisons
as well as within-population sequence comparisons from both populations 
should be included in the above likelihood
(where each pairwise comparison must be at a different, independent locus).
The above explicit formula for the likelihood allows rapid computation and 
maximization, so that ML estimates can easily be obtained, as well as AIC scores and 
likelihood ratios comparing the IIM model with
competing models such as the ``complete isolation" model and the symmetric
``isolation-with-migration" model.
Similar ML methods were first developed by Takahata et al.~(1995) 
for a number of different demographic models: a single population of constant 
size, a population undergoing an instantaneous change of size, and complete 
isolation models for two and for three species.
Innan and Watanabe~(2006) developed an extension of Takahata et al.'s 
MLE method to a more sophisticated model of gradual population 
divergence than the IIM model considered in this paper,
but their calculation of the likelihood relies on numerical computation 
of the probability density function of the coalescence 
time using recursion equations on a series of time points (and then 
numerically integrating over the 
coalescence time to find the probability of $k$ nucleotide differences), 
which can be time-consuming; in addition, the accuracy of their recursion and
likelihood calculation depend on the number of time points considered.
The method described in the present paper 
assumes a simpler model than Innan and Watanabe's, but is faster 
because both the calculation of the pdf of the coalescence time and the 
integration over the coalescence time to find the probability of $k$
nucleotide differences have already been done, in an exact way, 
to give equations~(\ref{distribution0}) and~(\ref{distribution1}) above,
leaving far less computation to be done.

In order to obtain good starting values for the likelihood maximization
under the IIM model, our implementation first fits a complete isolation  
model, as the latter model gives a more tractable likelihood surface and is
therefore less sensitive to the choice of starting values. The parameter 
estimates obtained for the complete isolation model are then used as starting 
values to fit the IIM model; this was found to reduce the possibility that 
the program might otherwise converge on a local maximum rather than on the 
global maximum of the likelihood under the IIM model.
Whilst the theoretical results obtained 
make it possible to directly compute ML estimates of the original parameters 
of the IIM model as described above ($\tau_1, \tau_0, c_1, c_2, a, M, \theta$), 
our implementation uses the following reparameterizations 
as this improved performance and robustness:
\begin{equation}
\begin{array}{l}
T_1 = \theta \tau_1,\,\,  
V  = \theta (\tau_0 - \tau_1), \vspace*{2mm} \\ 
\theta_1 = c_1 \theta, \,\,
\theta_2 = c_2 \theta, \,\,
\theta_a = a \theta 
\end{array}
\label{pars}
\end{equation}
(similar to the choice of parameters in, for example, Yang~2002, 
and Hey and Nielsen~2004),
i.e. $T_1$ and $V$ represent respectively the time since the end of 
gene flow and the duration of the period of gene flow, 
measured by twice the expected number of mutations per lineage during
the period concerned;
$\theta_1, \theta_2$ and $\theta_a$ are the ``population size parameters" 
of the current descendant populations and the ancestral population, 
respectively ($\theta=4N\mu$ is the population size parameter of each 
descendant population during the migration stage of the model).
ML estimates are obtained jointly for the parameters $T_1, V, \theta_1, 
\theta_2, \theta, \theta_a, M$, and these can readily be converted to
ML estimates of the original model parameters if required.
Our computer code is included as Supplementary Material.

Equations~(1) and~(2) rely on the assumption of symmetric migration and equal population sizes during the period of gene flow. Without these assumptions, an explicit analytical formula for the likelihood becomes difficult to obtain. Simulation results suggest however that our method is reasonably robust to minor violations of these assumptions;
furthermore, it is possible to extend our method to allow for asymmetric migration and unequal population sizes (Costa RJ and Wilkinson-Herbots HM, work in progress).

\section*{Results}

\subsection*{Application to Drosophila data}

To illustrate the MLE method for the IIM model described above, we applied 
this method to the genomic
data set of {\em D. simulans} and {\em D. melanogaster} compiled and analyzed 
by Wang and Hey~(2010), and reanalyzed by Lohse et al.~(2011), who both used
an IM model assuming a constant migration rate from the onset of speciation
until the present. 
The data consist of alignments of 30,247 blocks of intergenic sequence of 
500 bp each, from two inbred lines of {\em D. simulans} and from one inbred 
line each of 
{\em D. melanogaster} and {\em D. yakuba}, and have been pre-processed as 
described in Wang and Hey~(2010) and Lohse et al.~(2011). We also follow 
these authors in using {\em D. yakuba} as an outgroup to estimate the 
relative mutation rate at each locus. As our method uses the number 
of nucleotide differences between one pair of 
sequences at each locus, we had to choose two of the three sequences
(which we will denote by {\em D.sim}1, {\em D.sim}2 and {\em D.mel} for brevity)
at each locus. There are of course many ways in which this can 
be done (for example, one could choose two sequences at random at each locus,
as done by Wang and Hey~2010). In order to use all the data, 
and to be able to check to what extent our results depend on our choice of 
sequences, we took the following approach: three (overlapping) 
pairwise data sets were formed by alternately assigning loci to the 
comparisons {\em D.mel~-~D.sim}1,
{\em D.mel~-~D.sim}2 and {\em D.sim}1~-~{\em D.sim}2, where data set 1 starts with {\em D.mel}~-~{\em D.sim}1
at locus 1 ({\em D.mel}~-~{\em D.sim}2 at locus 2, {\em D.sim}1~-~{\em D.sim}2 at locus 3, and so on), 
data set 2 uses {\em D.mel}~-~{\em D.sim}2 at locus 1 ({\em D.sim}1~-~{\em D.sim}2 at locus 2, \ldots), 
and data set 3 starts with {\em D.sim}1~-~{\em D.sim}2 at locus 1 ({\em D.mel}~-~{\em D.sim}1 at locus 2, 
\ldots). 
Thus each sequence at each locus 
is used in exactly two of the three data sets, and each data set contains
between-species differences at approximately 20,000 loci and 
within-species ({\em D. simulans}) differences at approximately 10,000 loci.
ML estimates and estimated standard errors for the IIM model parameters  $(T_1, V, \theta_1, \theta, \theta_a, M)$\footnote{The population size parameter $\theta_1$ represents the current size of {\em D. simulans}. We used the main Wang and Hey (2010) data set used also by
Lohse et al.~(2011), which does not include any {\em D. melanogaster} pairs and hence contains no information on the population size parameter $\theta_2$ corresponding to the current size of 
{\em D. melanogaster} (this parameter does not appear in the likelihood of these data and thus cannot be estimated here).}
were obtained for each of the 
three data sets and then averaged over the three data sets. 
For comparison, in addition to fitting an IIM model as described above, 
we also obtained ML estimates assuming 
an IIM model with $\theta_1=\theta$ (or equivalently, $c_1=1$, i.e. 
not allowing for a change of population size at the end of the 
migration period), a symmetric IM model (which corresponds to putting $T_1=0$ 
in the IIM model), and a complete isolation model (which corresponds to 
assuming $M=0$ in addition to $T_1=0$; the size of descendant population~2 is irrelevant 
here). The results are given in Table~\ref{tab1}. 
\begin{table}[h]
\caption{ML estimates, maximized loglikelihood ($\ln\hat{L}$), likelihood ratio test statistic ($\Lambda$) and AIC score, 
obtained by fitting the different models to the three sets of Drosophila sequence data.} 
\vspace*{-1mm}
\begin{center}
{\scriptsize \begin{tabular}{llrrrrrrrrr} \hline \vspace*{-2mm} \\ 
 data & model & $\hat{T}_1$ \, & $\hat{V}$ \, & 
 $\hat{\theta}_1$ \, & $\hat{\theta}$ \, & $\hat{\theta}_a$ \, & $\hat{M}$ \,
 & ~~~~~~~~~~~~~~~$\ln\hat{L}$ & $\Lambda$ & AIC~ 
\\  \hline \vspace*{-1mm} \\
 set 1 & isolation &       &  13.66 \, &         & 5.67 \, & 4.69 \, &           & -90,097.17 & 1,175.66 & 180,200.34 \vspace{0.2cm}
\\ 
      & IM        &          & 14.79 \, &         & 5.54 \, & 3.97 \, & 0.0209 \, & -89,509.34 &   270.65 & 179,026.68 \vspace{0.2cm}
\\ 
      & IIM with $\theta_1=\theta$~~~~ 
                  &  5.87 \, &  9.99 \, &         & 5.47 \, & 3.48 \, & 0.1389 \, & -89,374.01 &   518.43 & 178,758.03 \vspace{0.2cm}
\\ 
      & IIM       &  7.22 \, &  9.38 \, & 6.70 \, & 2.68 \, & 3.22 \, & 0.0888 \, & -89,114.80 &          & 178,241.60   
\\ 
      & (s.e.)   & (0.20)   & (0.16)   & (0.11)  & (0.12)  & (0.09)  & (0.0059)  &            &          &            
\\ \hline  \vspace*{-1mm} \\
 set 2 & isolation &          & 13.57 \, &         & 5.72 \, & 4.77 \, &        \, & -90,339.14 & 1,185.38 & 180,684.28 \vspace{0.2cm}
\\ 
      & IM        &          & 14.80 \, &         & 5.58 \, & 3.98 \, & 0.0227 \, & -89,746.45 &   347.39 & 179,500.90 \vspace{0.2cm}
\\ 
      & IIM with $\theta_1=\theta$ 
                  &  6.27 \, &  9.88 \, &         & 5.49 \, & 3.37 \, & 0.1793 \, & -89,572.76 &   685.95 & 179,155.51 \vspace{0.2cm}
\\ 
      & IIM       &  7.62 \, &  9.43 \, & 6.85 \, & 2.31 \, & 3.05 \, & 0.0904 \, & -89,229.78 &          & 178,471.56   
\\
      & (s.e.)    & (0.17)   & (0.15)   & (0.11)  & (0.11)  & (0.09)  & (0.0054)  &            &          &            
\\ \hline  \vspace*{-1mm} \\
 set 3 & isolation &          & 13.63 \, &         & 5.59 \, & 4.72 \, &           & -89,979.56 & 1,094.44 & 179,965.12 \vspace{0.2cm}
\\ 
      & IM        &          & 14.69 \, &         & 5.48 \, & 4.04 \, & 0.0193 \, & -89,432.34 &   291.86 & 178,872.68 \vspace{0.2cm}
\\ 
      & IIM with $\theta_1=\theta$ 
                  &  6.29 \, &  9.67 \, &         & 5.39 \, & 3.46 \, & 0.1635 \, & -89,286.41 &   548.28 & 178,582.82 \vspace{0.2cm}
\\ 
      & IIM       &  7.31 \, &  9.31 \, & 6.62 \, & 2.56 \, & 3.23 \, & 0.0870 \, & -89,012.27 &          & 178,036.54 
\\
      & (s.e.)    & (0.20)   & (0.16)   & (0.11)  & (0.12)  & (0.09)  & (0.0058)  &            &          &            
\\ \hline  \vspace*{-1mm} \\
 average & IIM     &  7.39 \, &  9.38 \, & 6.72 \, & 2.52 \, & 3.16 \, & 0.0887 \, &            &          &      
\\ 
        & (s.e.)  & (0.19)   & (0.15)   & (0.11)  & (0.11)  & (0.09)  & (0.0057)  &            &          &            
\\ \hline  \vspace*{-4mm} \\
\end{tabular}}
\end{center}
{\footnotesize NOTE -- The values of the likelihood ratio test statistic $\Lambda$ shown are for the comparison of the model concerned (considered to be the null model) 
against the model immediately underneath it.
Estimated standard errors (numbers in brackets) are provided for the model with the best fit, i.e. the full IIM model.
The bottom section of the table gives the parameter estimates under the IIM 
model, averaged over the three data sets;
for each parameter, the average of the estimated standard errors for the 
three data 
sets is given in brackets, which serves as an estimated {\em upper bound} on 
the standard error of the averaged parameter estimate
(estimates of the exact standard errors would be hard to obtain as the 
three data sets overlap).}
\label{tab1}
\end{table}
Fitting the full IIM model took   
approximately 20 seconds for each of the three data sets, on a desktop PC. 
Estimated mutation rates, and hence the estimates of the parameters
$T_1, V, \theta_1, \theta$ and $\theta_a$ defined in~(\ref{pars}), are averages 
over all the loci considered. Mutation rate heterogeneity between loci was
accounted for by estimating the relative
mutation rate at each locus from comparison with 
the outgroup {\em D. yakuba},
as proposed in Wang and Hey~(2010) (see ``Materials and Methods" for further details); 
for simplicity these relative mutation 
rates are treated as known constants (see also Yang~1997, 2002), i.e. uncertainty about the relative 
mutation rates is ignored.
Table~\ref{tab2} shows the estimates (averages over the three data sets) 
converted to times in years, diploid effective population sizes, 
and the migration rate per generation, for each of the four models considered.
\begin{table}[!bhtp]
\caption{ML estimates of {\em D. simulans - D. melanogaster} divergence times, effective population 
sizes and migration rate per generation, under the different models 
considered, obtained with the full sequence data.} 
\vspace*{-1mm}
\begin{center}
{\small
\begin{tabular}{lcccccc} \hline \vspace*{-3mm} \\ 
model & $\hat{t}_1$ \, & $\hat{t}_0$ \, & 
$\hat{N}_1$ \, & $\hat{N}$ \, & $\hat{N}_a$ \, & $\hat{m}$ \\ \hline \vspace*{-2mm} \\ 
isolation      &                 & 2.97my          &               & 6.18m         & 5.16m         &                                         \vspace{0.2cm}     \\ 
IM               &                 & 3.22my          &               & 6.04m         & 4.36m         & ~$8.68 \times 10^{-10}$ \vspace{0.2cm}     \\ 
IIM with $c_1=1~~~~~$ & 1.34my          & 3.49my          &               & 5.95m         & 3.75m         & $6.75 \times 10^{-9}$ \vspace{0.2cm}    \\  
IIM              & 1.61my    & 3.66my    & 7.33m    & 2.75m   & 3.45m    & $8.11 \times 10^{-9}$  \\ 
(s.e.)           & (0.04my) & (0.03my) & (0.12m) & (0.12m) & (0.10m) & ($0.59 \times 10^{-9}$)~ \\ \hline \vspace*{-4mm} \\
\end{tabular}}
\end{center}
{\footnotesize NOTE -- The abbreviations ``my" and ``m" stand for ``million years" and 
``million individuals". The times $t_1$ and $t_0$ denote, respectively, the time since
complete isolation of {\em D. simulans} and {\em D. melanogaster} (for the IIM model), 
and the time since the onset of speciation (for all models), i.e. these are the times $\tau_1$ and 
$\tau_0$ converted into years (see Fig.1). $N$ denotes the effective size of {\em
D. simulans} in the isolation and IM models, and during the migration stage
of the IIM model; $N_1=c_1 N$ denotes the present effective size of {\em D. simulans}
in the full IIM model; $N_a = a N$ is the effective size of the ancestral population. 
The estimates shown are the averages of the estimates obtained
for the three (overlapping) data sets described in the text. For the IIM model, the 
averaged estimated standard error is also given (in brackets) for each 
parameter; this is an estimated upper bound on the standard error of the 
averaged parameter estimate.}  
\label{tab2}
\end{table}
These conversions assume a generation time of 0.1 year 
and a 10 million year speciation time between {\em D. yakuba} and 
{\em D. melanogaster}/{\em D. simulans}
(as assumed by Wang and Hey~2010, and by Lohse et al.~2011; 
see also Powell~1997).
It is seen that the IIM model places the onset of speciation 
($\hat{t}_0=3.66$ million years) further back into the past than both the isolation model
and the IM model, whereas the estimated time since complete isolation 
of the two species under the IIM model ($\hat{t}_1=1.61$ million years) is 
more recent than under the isolation model.
Under the IIM model we obtain an estimated ancestral effective population
size of $3.45$ million 
individuals, splitting into two populations containing $2.75$ million individuals 
each during the time interval when gene flow occurred, with {\em D. simulans}
expanding to a current effective population size of $7.33$ million individuals.
Note that the estimated migration rate per generation ($\hat{m}$) is an order of 
magnitude higher under the IIM model than under the IM model.

On the one hand, our averaging of the estimated standard errors over the three data sets (bottom rows of Tables~\ref{tab1} and~\ref{tab2}) will have given us an overestimate of
the standard error of the averaged parameter estimates.
On the other hand, however, it should be noted that these estimated standard errors underrepresent
the true total amount of uncertainty, as they do not account for uncertainty about the relative mutation rates at the different loci (which have been treated as known constants, whereas in practice we have estimated them 
from comparison with {\em D. yakuba}).
Care should be taken therefore in the interpretation of the estimated standard errors stated in Tables~\ref{tab1} and~\ref{tab2}.

Table~\ref{tab1} also gives the maximized loglikelihood ($\ln\hat{L}$)
for the different models fitted, for each of the three data sets.
It is seen that, amongst the models being compared, the full IIM model
consistently gives by far the best loglikelihood value, though of course 
this model also has the largest number of parameters. 
The easiest way of comparing the fit of the different models is by 
using Akaike's Information Criterion, AIC, which was designed to compare
competing models with different numbers of parameters (AIC scores were 
also used in, for example, Takahata et al.~1995, Nielsen and Wakeley~2001, 
and Carstens et al.~2009). 
For each model (and for the same data), AIC is defined as
$$
\mbox{AIC} = -2 \ln\hat{L} + 2 \times \mbox{number of 
independently adjusted parameters within the model}
$$
(Akaike~1972, 1974). 
Thus a larger maximized likelihood leads to a lower AIC score, 
subject to a penalty for each additional model parameter.
The ``Minimum AIC Estimate" (MAICE) 
is then defined by the model (and the 
Maximum Likelihood estimates of the model parameters) which gives the 
smallest AIC value 
amongst the competing models considered.
Table~\ref{tab1} includes, for each data set, the AIC scores of the different models 
considered.
It is seen that for each of the three data sets, the MAICE is given
by the full IIM model. Thus, out of all the models considered here, the 
IIM model provides the best fit to the data, as measured by the AIC scores.
The results in Table~\ref{tab1} also show that an IIM model with $\theta_1=\theta$ 
has a substantially better AIC score than a symmetric IM model, suggesting 
that the improved fit of the IIM model compared to the IM model is indeed
due at least in part to accounting for the eventual complete isolation of
the two species, and is not due solely to allowing differences between 
population sizes.

An alternative approach is to perform a series of likelihood ratio tests 
for nested models (see, for example, Cox~2006, Section 6.5 on ``Tests and
model reduction").
Focusing on any one of the three data sets in Table~\ref{tab1}, if we start with 
the full IIM model and look upwards, each of the models considered 
reduces to the model immediately above it by fixing the value of one parameter:
respectively $\theta_1=\theta$, $T_1=0$, and $M=0$.
Each pair of ``neighbouring" models can then be formally compared by means
of a likelihood ratio test, where the null hypothesis represents the 
simpler of the two models (i.e. the one with fewer free parameters). 
If the null hypothesis is true, then the test statistic
$$\Lambda=2 \times \left(\ln\hat{L}(\mbox{alternative model})-\ln\hat{L}(\mbox{null 
model}) \right)$$
should approximately follow a $\chi^2_1$ distribution 
if the ``null" value of the parameter concerned is an interior point
of the parameter space (as is the case when we test $\theta_1=\theta$),
whereas we might
expect $\Lambda$ to follow approximately a $\frac{1}{2} \times 0 + \frac{1}{2} \times \chi^2_1$ mixture if 
the ``null" value of the parameter concerned lies on the boundary of the 
parameter space (as is the case in the tests of $T_1=0$ and $M=0$); 
using a $\chi^2_1$ distribution in the latter case is conservative
(Self and Liang~1987; see also our simulation results below).
The value of the likelihood ratio test statistic $\Lambda$ is given in 
Table~\ref{tab1} for each possible null model,
when evaluated against the model immediately underneath it.
For each model comparison, the value of the test statistic $\Lambda$ is so large that the   
simpler model is rejected in favour of the more complex model underneath it,
with, in each case, a very small p-value ($p << 10^{-10}$) providing 
overwhelming evidence against the simpler model. Thus these likelihood ratio tests also identify the full IIM model 
as the most plausible amongst the different models considered.

\subsection*{Simulation results}

Simulations were done to assess how well our method performs and how 
reliable the estimates obtained in the previous section are.
From the three pairwise sets of Drosophila sequence data considered in Table~\ref{tab1}, 
we arbitrarily selected the first one (``data set~1") and mimicked this
data set in our simulations. 

\subsubsection*{Simulations from the IIM model}
\begin{figure}[!b]
\centering
\vspace*{-2.5cm}
\includegraphics[width=15.2cm, angle=0]{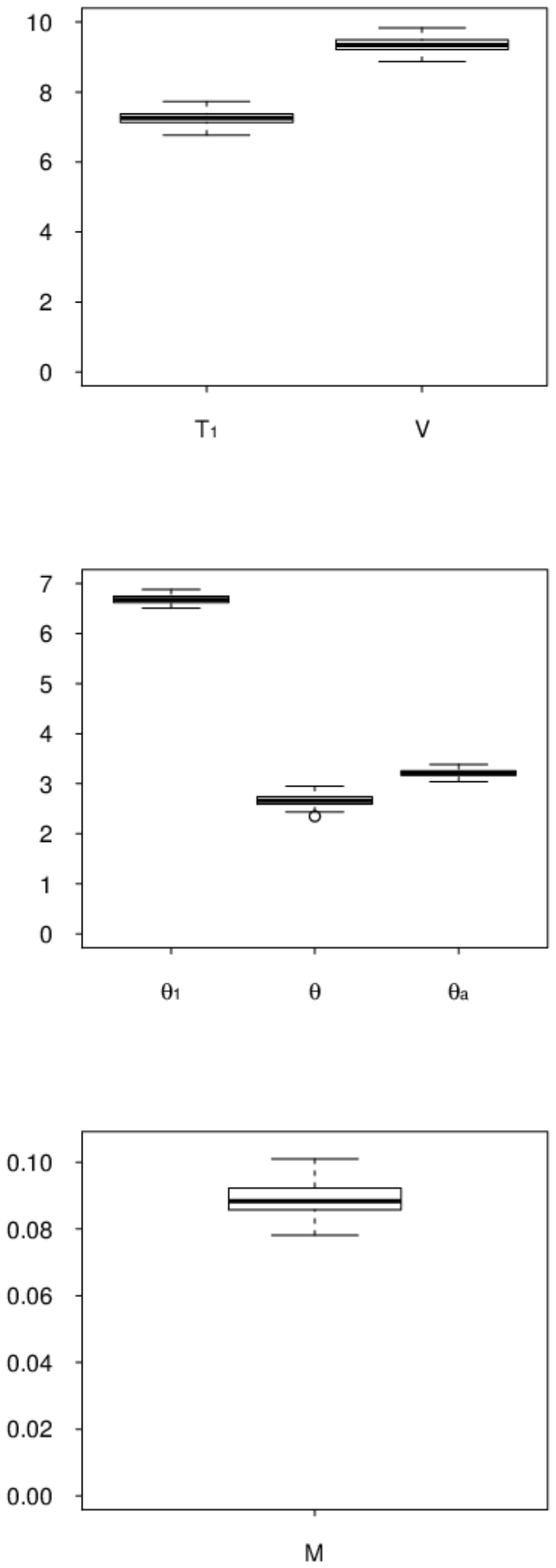}
\vspace*{-2.5cm}
\caption{ML estimates obtained for 100 simulated data sets. 
The simulations assumed an IIM model with true parameter values
$T_1=7.22$, $V=9.38$, $\theta_1=6.70$, $\theta=2.68$, $\theta_a=3.22$ 
and $M=0.0888$. Each simulated data set consists of the number of nucleotide 
differences between two DNA sequences from descendant population~1 at 
10,082 independent loci, and between two DNA sequences from different 
descendant populations at 20,165 independent loci. 
Different loci were assumed to have different mutation rates: the relative mutation 
rates used at the different loci match those of the Drosophila data considered in this paper.}
\label{fig2}
\end{figure}

Each simulated data set consists of
the number of nucleotide differences between a pair of DNA sequences sampled 
from descendant population~1 for 10,082 independent loci, and the number of differences between 
a pair of DNA sequences taken from different descendant populations 
for 20,165 independent loci, generated under the full
IIM model with ``true" parameter values equal to the estimates obtained for 
data set~1 (i.e. $T_1=7.22$, $V=9.38$, $\theta_1=6.70$, $\theta=2.68$,
$\theta_a=3.22$ and $M=0.0888$) and the infinite sites model of neutral 
mutation.  These numbers of loci for 
both types of comparison match those in Drosophila data set~1, and 
so do the relative mutation rates assumed at the different loci.
One hundred such data sets were generated. For each simulated data set,
ML estimates of the parameters were obtained using our R program for the 
IIM model, i.e. the method described in ``New Approaches". The resulting estimates are shown in Figure~\ref{fig2}.
As one would expect, it is seen that (at least for a large sample size) our ML estimation procedure gives estimates centred on, and close to, the ``true" parameter values.

In order to assess whether our method enables us to correctly select the IIM model from amongst the different models considered, when the IIM model is in fact the true underlying model, we also mimicked 
the type of analysis as was shown for data set~1 in Table~\ref{tab1}: for each simulated data set, we fitted an isolation model, a symmetric IM model, an IIM model with $\theta_1=\theta$, and a full IIM model, and computed the AIC scores and the values of the likelihood ratio test statistic $\Lambda$. Both procedures (using AIC scores or likelihood ratio tests) correctly identified the full IIM model as the best-fitting model, for each of the 100 simulated data sets. The smallest difference obtained between the AIC scores of any two neighbouring models was in fact $242.96$, with the simpler model always having the 
worse AIC score, and the smallest difference observed between the AIC scores of the full IIM model 
and any other model
was $409.93$, indicating that the IIM model could be identified with ease. 
For the likelihood ratio approach, comparing the value of the test statistic $\Lambda$ for a pair of neighbouring models 
with the $\chi^2_1$ distribution
gave in all cases a $p$-value much smaller than $10^{-10}$,
providing extremely strong evidence 
against the simpler of the two models compared
(the smallest value of $\Lambda$ obtained for any pair of neighbouring models was $244.96$, 
whereas the $(1-10^{-10})$ quantile of the $\chi^2_1$ distribution is only $41.82$).
When comparing the full IIM model directly with the isolation model by means of a likelihood ratio test, the smallest value of $\Lambda$ observed 
amongst the simulated data sets 
was $1158.40$, again giving a $p$-value very much smaller than $10^{-10}$ (using the $\chi^2_3$ distribution) for each of the 100 simulated data sets.

\subsubsection*{Simulations from the isolation model}
We also simulated 100 data sets under the isolation model, to further investigate the performance
of our method in identifying the correct model, and whether false positives may be produced -- in particular, whether a signal of gene flow
may be obtained when in reality there was no gene flow. The ``true" parameter values assumed in the simulations were those obtained from fitting an isolation model to Drosophila data set~1 (see Table~\ref{tab1}). Two of the simulated data sets were problematic, in that only the isolation model could be fitted: due to some large values ($> 200$) for the simulated numbers of nucleotide differences between pairs of sequences from different species, R was unable to evaluate the likelihood for any of the other models in the relevant part of the parameter space.
For the remaining $98$ simulated data sets for which all models could be fitted, a likelihood ratio test using as the null distribution the naive 
$\chi^2$ distribution (with degrees of freedom equal to the difference between the number of parameters in the two models being compared) gave better results than did comparison of AIC scores.
On the basis of AIC scores, an incorrect model was selected for as many as 19 of the simulated data sets: in 6 cases the IM model gave the lowest AIC score, in 2 cases the IIM model with $\theta_1=\theta$, and in 11 cases the full IIM model. However, in 6 of the 11 cases where the full IIM model was selected, the estimated migration rate was $0$ so that this ``IIM" model 
was in fact an isolation model but with an additional small change of population size. 

\begin{figure}[!bh]
\centering
\vspace*{-1.3cm}
\includegraphics[width=8.4cm, angle=0]{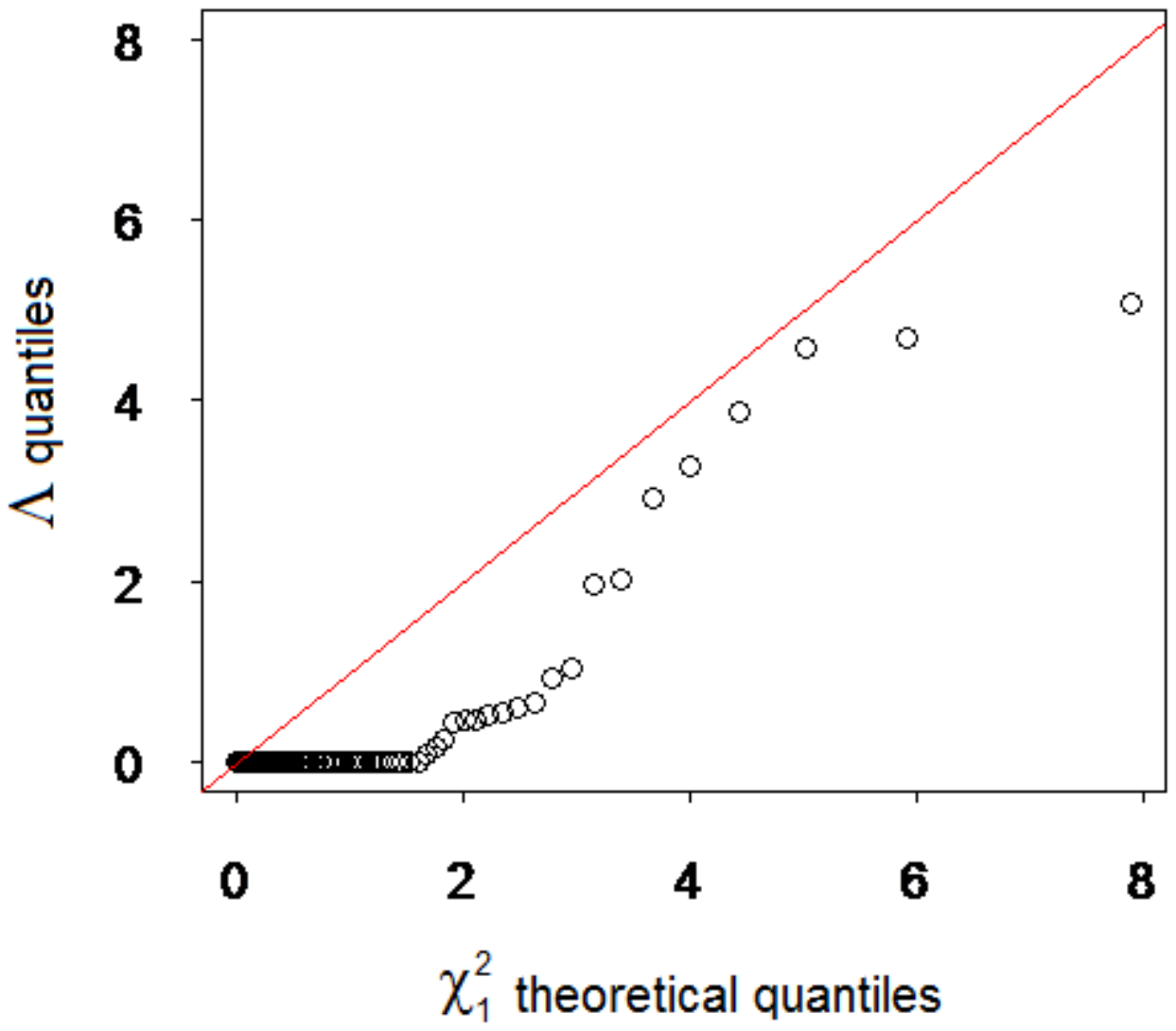}
\includegraphics[width=8.4cm, angle=0]{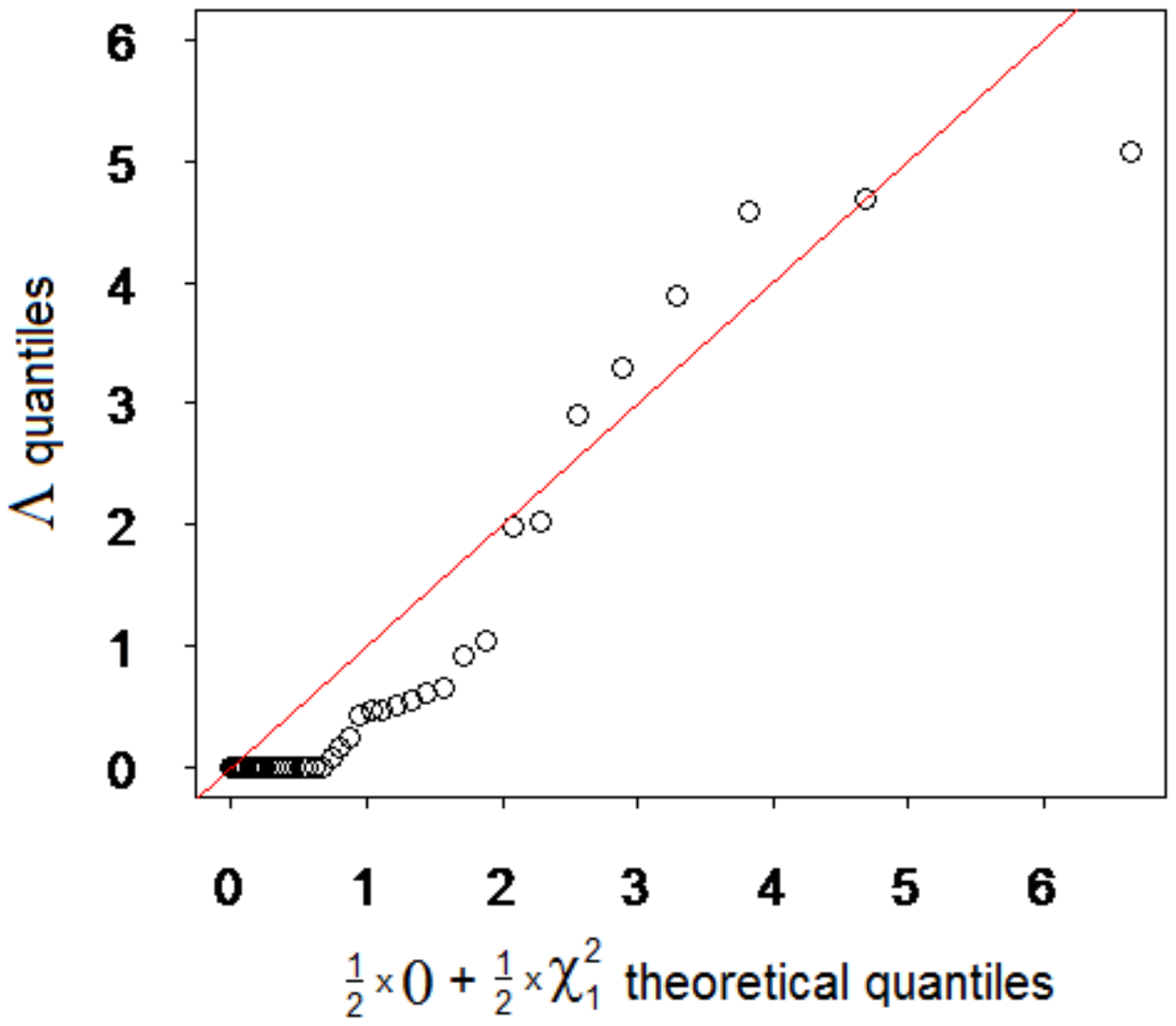}
\vspace*{-5.9cm}
\caption{QQ-plot of the likelihood ratio test statistic $\Lambda=2 \times (\ln\hat{L}(\mbox{IM})-\ln\hat{L}(\mbox{isolation}) )$ against the $\chi^2_1$ distribution (left), and against the mixture $\frac{1}{2} \times 0 + \frac{1}{2} \times \chi^2_1$ (right).
On the vertical axis we have plotted the sample quantiles of the values of $\Lambda$ observed when testing  the isolation model against the IM model for $100$ data sets simulated from the isolation model 
(2 data sets have been omitted as R was unable to fit the IM model - please see the main text for further details). The line $y=x$ is also shown for ease of comparison.}
\label{fig3}
\end{figure}

A likelihood ratio test of the isolation model against the IM model at a significance level of $1\%$ resulted in acceptance of the isolation model for all 98 data sets, 
regardless of whether we used $\chi^2_1$ or $(\frac{1}{2} \times 0 + \frac{1}{2} \times \chi^2_1)$ as the null distribution. At a significance level of $5\%$, the isolation model was rejected for 4 of the 98 simulated data sets if we used $\chi^2_1$, and for 6 of the simulated data sets if the mixed $\chi^2$ distribution 
was used. Figure~\ref{fig3}
 shows that the use of $\chi^2_1$ as the null distribution is indeed conservative, and suggests that this may be preferable to using the mixed $\chi^2$ distribution.

\begin{figure}[!t]
\centering
\vspace*{-1.2cm}
\includegraphics[width=8.4cm, angle=0]{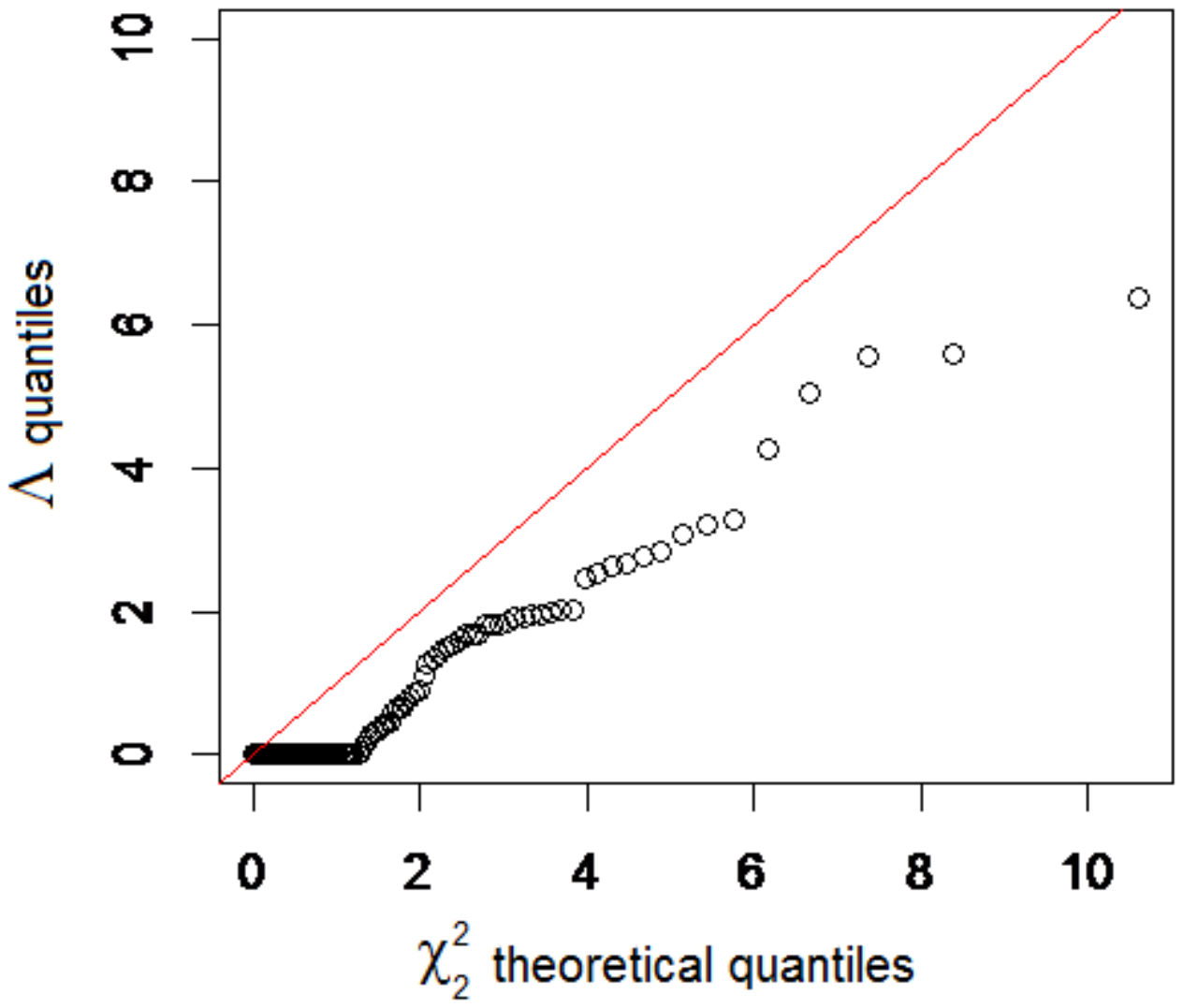}
\includegraphics[width=8.4cm, angle=0]{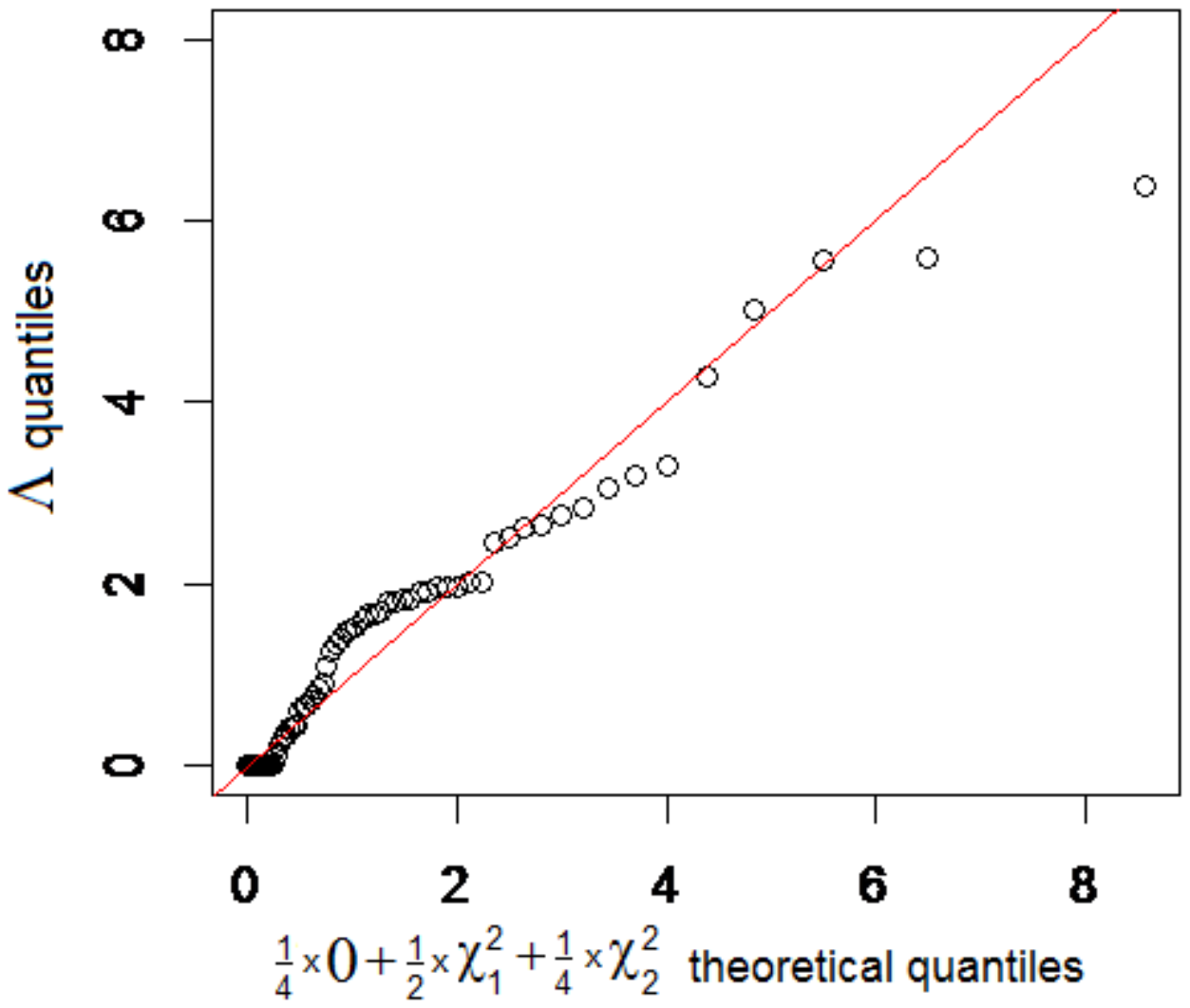}
\vspace*{-5.9cm}
\caption{QQ-plot of the likelihood ratio test statistic $\Lambda=2 \times (\ln\hat{L}(\mbox{IIM with $\theta_1=\theta$})-\ln\hat{L}(\mbox{isolation}) )$ against the $\chi^2_2$ distribution (left), and against the mixture $\frac{1}{4} \times 0 + \frac{1}{2} \times \chi^2_1 + \frac{1}{4} \times \chi^2_2$ (right).
On the vertical axis we have plotted the sample quantiles of the values of $\Lambda$ observed when testing  the isolation model against the IIM model with $\theta_1=\theta$ for $100$ data sets simulated from the isolation model 
(2 data sets have been omitted as R was unable to fit the IIM model with $\theta_1=\theta$).
The line $y=x$ is also shown for ease of comparison.}
\label{fig4}
\end{figure}

Similarly, at a significance level of $1\%$, a likelihood ratio test of the isolation model against the IIM model with $\theta_1=\theta$ led to acceptance of the isolation model for all 98 data sets, whether we used $\chi^2_2$ or the mixture $(\frac{1}{4} \times 0 + \frac{1}{2} \times \chi^2_1 + \frac{1}{4} \times \chi^2_2)$ as the null distribution. At a significance level of $5\%$, the use of 
$\chi^2_2$ resulted in rejection of the isolation model for 1 of the 98 data sets, whilst the use of $(\frac{1}{4} \times 0 + \frac{1}{2} \times \chi^2_1 + \frac{1}{4} \times \chi^2_2)$ as the 
null distribution led to rejection of the isolation model for 5 of the data sets. The QQ-plots shown in Figure~\ref{fig4} confirm again that the use of $\chi^2_2$ in this case is conservative. 

\begin{figure}[!h]
\centering
\vspace*{-1.3cm}
\includegraphics[width=8.4cm, angle=0]{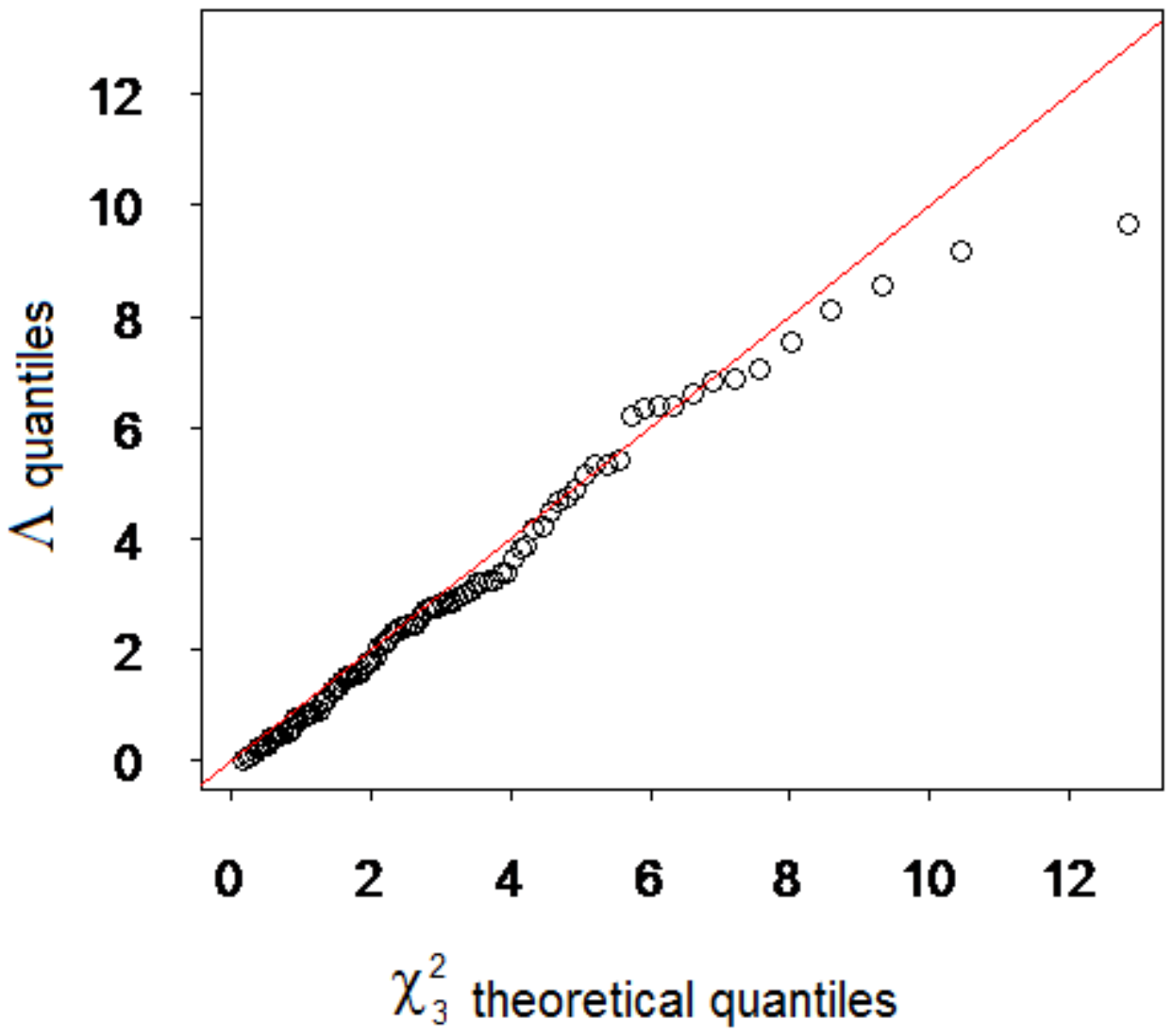}
\includegraphics[width=8.4cm, angle=0]{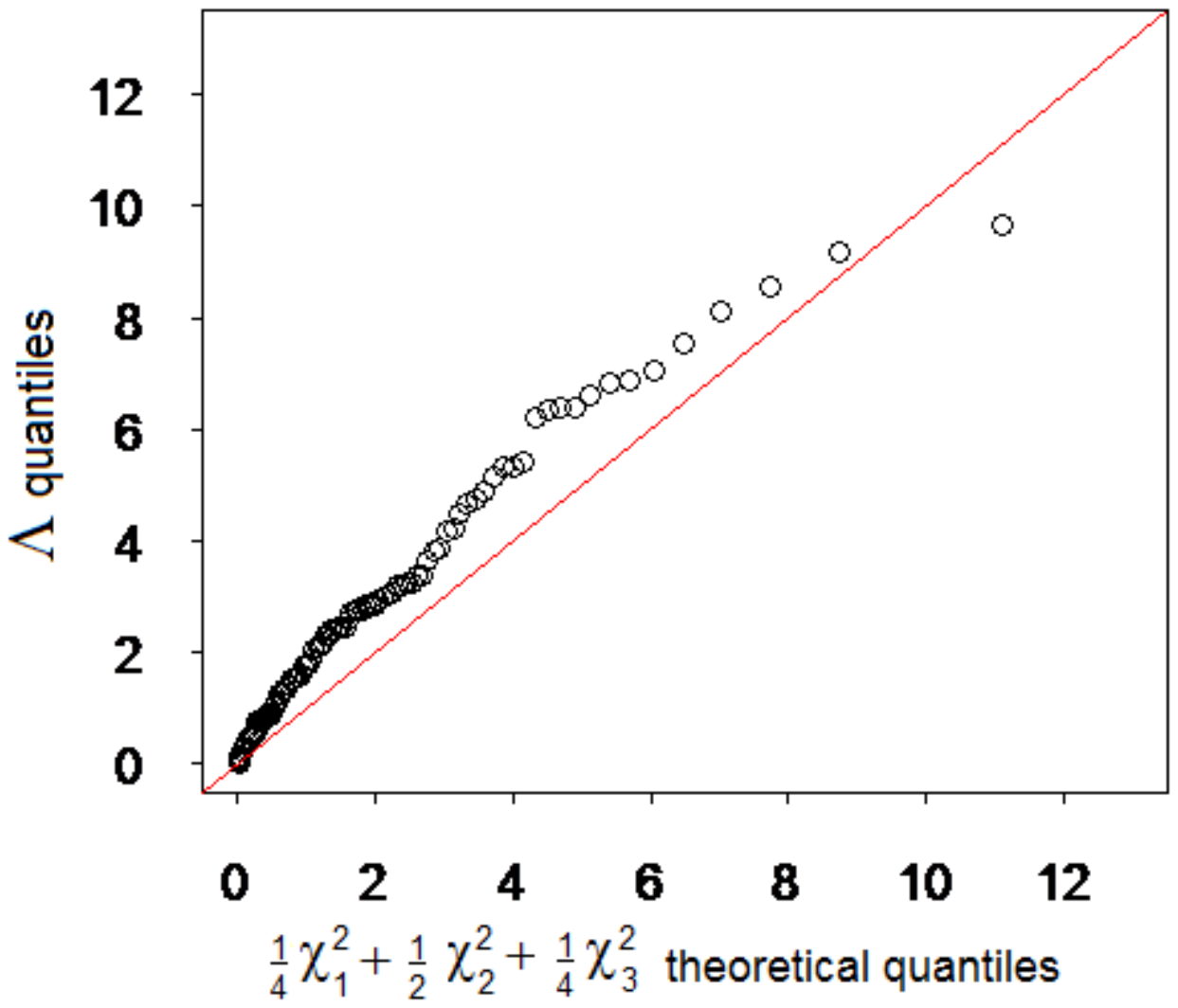}
\vspace*{-5.9cm}
\caption{QQ-plot of the likelihood ratio test statistic $\Lambda=2 \times (\ln\hat{L}(\mbox{IIM})-\ln\hat{L}(\mbox{isolation}) )$ against the $\chi^2_3$ distribution (left), and against the mixture $\frac{1}{4} \times \chi^2_1 + \frac{1}{2} \times \chi^2_2 + \frac{1}{4} \times \chi^2_3$ (right).
On the vertical axis we have plotted the sample quantiles of the values of $\Lambda$ obtained when testing the isolation model against the full IIM model for $100$ data sets simulated from the isolation model 
(2 data sets have been omitted as R was unable to fit the IIM model).
The line $y=x$ is also shown for ease of comparison.}
\label{fig5}
\end{figure}

Similarly, the QQ-plots in Figure~\ref{fig5} indicate that the use of the $\chi^2_3$ distribution is conservative when testing the isolation model against the full IIM model, and that this may be preferable to the use of a mixed $\chi^2$ distribution.
A likelihood ratio test at a significance level of $1\%$
led to acceptance of the isolation model for all 98 data sets if a $\chi^2_3$ distribution was used, and resulted in rejection of the isolation model for 1 of the 98 data sets when using the mixed
$(\frac{1}{4} \times \chi^2_1 + \frac{1}{2} \times \chi^2_2 + \frac{1}{4} \times \chi^2_3)$ distribution. At a significance level of $5\%$, using the $\chi^2_3$ distribution resulted in rejection of the isolation model for 4 of the 98 data sets (in 2 of these 4 cases the estimated migration rate was $0$, reducing the IIM model to an isolation model with an additional slight change of population size), whereas the use of the mixed $\chi^2$ distribution 
led to rejection of the isolation model for 12 of the 98 data sets (in 6 of these 12 cases, the estimated migration rate was $0$).
 
\section*{Discussion}

In this paper we have presented a very fast method to obtain ML estimates and to distinguish between different evolutionary scenarios, using nucleotide difference data from pairs of sequences at a large number of independent loci. The IIM model considered allows for an initial period of gene flow between two diverging populations or species, followed by a period of complete isolation, and is more appropriate in the context of speciation than the IM model commonly used in the literature (which assumes that gene flow continues at a constant rate all the way until the present time). We have illustrated the speed and power of our method by applying it to a large data set from two related species of Drosophila, and to simulated data. Fitting an IIM model to a data set of approximately 30,000 loci (with an average length of about 400 bp and varying mutation rates) from two species of Drosophila took approximately 20 seconds on a desktop PC; this time included fitting an isolation model first in order to obtain good starting values for the parameters.
Moreover, the results made it possible to distinguish between more and less plausible models (representing alternative evolutionary scenarios) with ease, and identified the IIM model as providing the best fit amongst the models considered.

The Drosophila data set studied in this paper was previously analysed by Wang and Hey~(2010), who compiled the data, and by Lohse et al.~(2011). Both fitted IM models to these data, assuming that gene flow occurred at a constant rate from the time of separation of the two species until the present time. In Tables~\ref{tab1} and~\ref{tab2}, alongside our results for the IIM model, we also included parameter estimates for a symmetric IM model, for the sake of comparison. As expected, our IM results are in close agreement with those of Lohse et al.~(2011)\footnote{See the results in Lohse et al.~(2011) for the full sequences rather than those for their trimmed version of the data. Note however that their treatment of mutation rate heterogeneity is slightly different from ours: in order to speed up the likelihood computation they grouped loci into 10 bins according to their outgroup divergence, whereas we have used the precise outgroup divergence for each locus. Another difference is their assumption that all population sizes are equal, including that of the ancestral population.}, who fitted two versions of the IM model, one with symmetric migration and one with migration in only one direction ({\em D. simulans} to {\em D. melanogaster} forward in time, motivated by Wang and Hey's results). They found that their estimated migration rate for the symmetric model is approximately half that obtained for the model with migration in one direction, i.e. both models gave the same estimated total number of migrants per generation between the two species in the two directions combined, whereas their estimates of the other parameters are virtually identical for the two models
(see p.23 of their Supporting Information). Our IM results are also in good agreement with those of Wang and Hey~(2010), even though they fitted a more general version of the IM model (allowing for unequal sizes of the two species and unequal migration rates in the two directions) and assumed the Jukes-Cantor model of mutation, whereas our results and those of Lohse et al.~(2011) assume the infinite sites model for its mathematical simplicity. Indeed, Lohse et al.~(2011) comment on how little difference the assumption of the infinite sites model makes to the IM results for the pairwise Drosophila data, compared to Wang and Hey's results based on the Jukes-Cantor model.
Of more interest is the comparison of these authors' IM results with those obtained for our IIM model.
We note firstly that their estimates of 3.04 million years (Wang and Hey~2010) or 2.98 million years (Lohse et al.~2011) for the 
 {\em simulans-melanogaster} speciation time, and also our IM estimate of 3.22 million years,  
fall in between our IIM estimates of 3.66 million years for the time since the onset of speciation ($t_0$) and 1.61 million years for the time since complete isolation of these two species ($t_1$).
Secondly, under the IM model the amount of gene flow between the two species (in the two directions combined) was estimated at 0.0134 migrant gene copies per generation (Wang and Hey~2010), 0.0255 migrant gene copies per generation (Lohse et al.~2011, asymmetric IM model), 0.0256 (Lohse et al.~2011, symmetric IM model), or 0.0210 migrant gene copies per generation (our IM results in Table~\ref{tab1}, averaged over the three data sets). These estimates are considerably smaller than the corresponding estimate
of 0.0887 migrant gene copies per generation during the period of migration under our IIM model.

Lohse et al.~(2011) also considered a trimmed version of the Wang and Hey (2010)  Drosophila data, where they shortened all loci to a fixed number of nucleotide differences between {\em D. melanogaster} and the {\em D. yakuba} outgroup, so that the estimated mutation rate for all loci was equal. This assumption that all loci have the same mutation rate massively speeds up the computation of the likelihood, since in this case the probability of $k$ differences needs to be calculated only once for each observed value of $k$, rather than having to calculate this probability over and over again for different mutation rates at different loci. Further to this idea, we also implemented a simplified, much faster, version of our MLE method to be used for data sets where all loci have the same mutation rate (this R code is also included as Supplementary Material). Applying this version of our IIM program to the trimmed Drosophila data (still approximately 30,000 loci) gave virtually instant results 
on a desktop PC, i.e. the computing time was $<< 1$ second. However, the substantial shortening of loci 
did result in loss of information, and the distinction between more and less plausible models was not always as clear anymore, as the differences between the values of the maximized loglikelihood for the different models (and hence the values of the likelihood ratio test statistic $\Lambda$, and the differences between the AIC scores of different models) were not as large as they were for the full sequence data. Nevertheless, the full IIM model could still be identified as giving the best fit amongst the models considered in this paper, both on the basis of AIC scores and by means of likelihood ratio tests. 
Table~\ref{tab3} gives the estimates obtained by fitting the IIM model to the trimmed data;
the conversion to times in years, diploid effective population sizes and migration rate per generation was done using a generation time of 0.1 year and a 10 million year 
speciation time between {\em D. yakuba} and \{{\em D. melanogaster}, {\em D. simulans}\}, as before.
The results are broadly in agreement with the corresponding estimates for the full sequence data (see Table~\ref{tab2}, bottom row).
\begin{table}[!tb]
\caption{ML estimates of {\em D. simulans - D. melanogaster} divergence times, effective population 
sizes and migration rate per generation, under the full IIM model, obtained with the trimmed sequence data.} 
\vspace*{-1mm}
\begin{center}
{\small \begin{tabular}{lcccccc} \hline  \vspace*{-3mm} \\ 
model~~~~~~~~~~ & $\hat{t}_1$ \, & $\hat{t}_0$ \, & 
$\hat{N}_1$ \, & $\hat{N}$ \, & $\hat{N}_a$ \, & $\hat{m}$ \\ \hline \vspace*{-2mm} \\ 
IIM      & 1.52my & 3.67my & 6.04m & 3.94m & 3.82m & $11.27 \times 10^{-9}$ \\ 
(s.e.)   & (0.16my) & (0.10my) & (0.11m) & (0.33m) & (0.18m) & ($2.89 \times 10^{-9}$) \\ \hline \vspace*{-4mm} \\
\end{tabular}}
\end{center}
{\footnotesize NOTE -- The abbreviations ``my" and ``m" stand for ``million years" and 
``million individuals". The times $t_1$ and $t_0$ denote, respectively, the time since
complete isolation of {\em D. simulans} and {\em D. melanogaster}, 
and the time since the onset of speciation, i.e. these are the times $\tau_1$ and 
$\tau_0$ converted into years (see Fig.1). The effective population sizes $N_1$, $N$ and $N_a$ refer respectively to the present {\em D. simulans} population, 
each population during the migration stage of the model, and the ancestral population. 
The estimates shown have been averaged over the three overlapping data sets described in the text. The 
averaged estimated standard error is also given (in brackets) for each 
parameter; this is an estimated upper bound on the standard error of the 
averaged parameter estimate.  }
\label{tab3}
\end{table}

The purpose of our analysis of the Drosophila data was merely to illustrate the potential of our method, rather than to draw any firm conclusions about the evolutionary history of the particular species concerned. Whilst the IIM model provides the best fit to the Drosophila data amongst the different models considered in this paper, it is likely
that more realistic models can be constructed which provide yet a better fit than the version of the IIM model considered here. 
Our assumption of equal population sizes and symmetric migration during the period of gene flow in the model is obviously an unrealistic oversimplification, but was made for the sake of mathematical tractability and computational speed. An extension of our method to a more general IIM model allowing for unequal migration rates and unequal population sizes during the migration stage of the model is possible and is being developed
(Costa RJ and Wilkinson-Herbots HM, work in progress).
Similarly, the current implementation of our method assumes the infinite sites model of neutral mutation because of its mathematical ease. Extensions of our method to other neutral mutation models (for example, the Jukes-Cantor model) can be done but have not yet been implemented. It should also be feasible to extend our method to incorporate more than two species. 

Whilst our method explicitly allows for mutation rate heterogeneity between loci (whether due to variation in sequence length or variation in the mutation rate per site, or both), our current implementation 
assumes that accurate estimates of the relative mutation rates of the different loci are available
(our implementation treats the relative mutation rates as known constants). This limitation will become less important as more and more genome sequences become available, allowing more accurate estimation of the relative mutation rates. For our analysis of the Drosophila data presented in this paper, however, the relative mutation rates of the different loci were estimated by comparison of \{{\em D. simulans}, {\em D. melanogaster}\} with the outgroup {\em D. yakuba}, as was done also by Wang and Hey~(2010) and by Lohse~et al.~(2011). 
Using more than one outgroup sequence (if available) to estimate the relative mutation rate at each locus should improve the accuracy.
It may also be possible to adapt our method to incorporate uncertainty about the relative mutation rates 
by modelling mutation rate variation as a random variable and integrating out over it (see Yang~1997, or the 
``all-rate" method proposed by Wang and Hey~2010).

\section*{Materials and Methods}

The Drosophila data considered in this paper are the {\em D. melanogaster - D. simulans} divergence data compiled and analyzed by 
Wang and Hey~(2010); we used the subset of the data that was studied also by Lohse et al.~(2011). The data consist of alignments of 
30,247 segments of intergenic sequence of length 500 bp each,  from two inbred lines of {\em D. simulans} and one inbred line of 
{\em D. melanogaster}, and from an inbred line of {\em D. yakuba} for use as an outgroup. The data have been pre-processed as described 
in Wang and Hey~(2010) and Lohse et al.~(2011); the version of the data used is that labelled ``WangHeyRaw" in the Supporting Information of Lohse et al.~(2011).
As our method uses pairwise nucleotide differences, subsets of the data were formed by selecting one pair of sequences from each locus (i.e. {\em D.mel}~-~{\em D.sim}1, {\em D.mel}~-~{\em D.sim}2, or {\em D.sim}1~-~{\em D.sim}2). Three such subsets were formed as described in the Section on ``Results"; these three data sets overlap as each sequence at each locus is used in two of the three data sets.
Parameter estimates and estimated standard errors were obtained for each of these three data sets separately and then averaged over the three data sets. This gives too large estimates of the standard errors of the averaged parameter estimates; however, due to the overlap between the data sets, more accurate estimated standard errors would be difficult to obtain.

Following Wang and Hey~(2010) and Lohse et al.~(2011), mutation rate heterogeneity between loci was accounted for by comparing the {\em D. melanogaster} and {\em D. simulans} sequences with the outgroup {\em D. yakuba}. 
Calculating the outgroup divergence
$d_l$ between \{{\em D. simulans}, {\em D. melanogaster}\} and {\em D. yakuba} at locus $l$ as a weighted average over the available sequences, assigning 25\% weight to each of the {\em simulans} sequences and 50\% weight to the single {\em melanogaster} sequence, 
the relative mutation rate $r_l$ of locus $l$ was estimated as $r_l=d_l/\bar{d}$,
where $\bar{d}$ is the average of the $d_l$ over all loci.
The scaled mutation rate at locus $l$ is then given by $\theta^{(l)} = r_l \theta$, where $\theta$ is the average scaled mutation rate over all the loci considered. The relative mutation rates $r_l$ are treated as fixed constants in the likelihood maximization (see also Yang~1997, 2002).   

Estimates of the parameters  
of the IIM model were obtained by maximizing the likelihood, given by equation~(\ref{likelihood}), using the reparameterization~(\ref{pars}).
In order to obtain reasonable starting values for the likelihood maximization under the IIM model, it can be helpful to first fit an isolation model, as the latter has a more tractable likelihood surface. Estimated standard errors are obtained from the inverted Hessian matrix. The computer code was written in R and is included as Supplementary Material.
Table~\ref{tab1} shows the results obtained for the three sets of Drosophila data. 

To convert the parameter estimates into readily interpretable units,
we followed Wang and Hey~(2010) and Lohse et al.~(2011) in using a generation time of $g=0.1$ year and a 10 million year speciation time between {\em D. yakuba} and \{{\em D. simulans}, {\em D. melanogaster}\} (see also Powell~1997), 
which gives us an estimate of $\bar{\mu} = 2.29 \times 10^{-7}$ for the mutation rate per locus per generation, averaged 
over all the loci included in the analysis.
The converted estimates given in Table~\ref{tab2} were then calculated according to the following equations:
\begin{eqnarray*}
t_1 & = & T_1 \times \frac{g}{2\bar{\mu}}\\
t_0 & = & (T_1 + V) \times \frac{g}{2\bar{\mu}} 
\end{eqnarray*}
for the times in years since complete isolation and since the onset of speciation, respectively;
$$N = \frac{\theta}{4 \bar{\mu}}$$
for the effective size (number of diploid individuals) of either population during the migration stage of the model, and similarly for all other population sizes; and
$$
m =  \frac{M}{\theta} \times \bar{\mu}
$$
for the migration rate per generation. Estimated standard errors for the converted estimates
were obtained by re-running a reparameterized version of the program, in terms of $T_0=T_1+V$ and $\Psi = \frac{M}{\theta}$ instead of $V$ and $M$, using the ML estimates already obtained as starting values, allowing easy computation of the Hessian matrix. As before, estimated standard errors were computed separately for each of the three data sets and then averaged. 
    
Following Lohse et al.~(2011) we also analyzed a trimmed version of the Wang and Hey~(2010) Drosophila data, where each locus was cut after 16 nucleotide differences between {\em D. melanogaster} and {\em D. yakuba}, and the remainder of the locus was ignored; the 2,090 loci which had fewer than 16 differences between {\em D. melanogaster} and {\em D. yakuba} were omitted altogether, leaving 28,157 trimmed loci for analysis. This trimming amounts to a shortening of the loci by roughly a factor of 3 on average. Again, three (overlapping) subsets of the data were formed, each containing one pair of sequences from each locus (i.e. {\em D.mel}~-~{\em D.sim}1, {\em D.mel}~-~{\em D.sim}2, or {\em D.sim}1~-~{\em D.sim}2) 
as described above and in the Section on ``Results". 
Parameter estimates and estimated standard errors were obtained for each of these three data sets separately and then averaged over the three data sets. We used a simplified version of our R code written specifically for data sets where all loci have the same estimated mutation rate; this simplified code is also included as Supplementary Material. The likelihood calculation uses the frequencies of the observed numbers of pairwise differences, which is much faster than the corresponding code for the case where different loci have different mutation rates: for the trimmed Drosophila data, instead of evaluating 28,157 terms in the loglikelihood (one term for each locus), only about 50 different terms need to be calculated (corresponding to the different values of the number of pairwise differences observed, within and between species) and multiplied by their frequencies, which hugely speeds up the likelihood computation and maximization. Table~3 shows the results obtained under the IIM model, after conversion into conventional units as explained above, using a mutation rate of $\mu=8 \times 10^{-8}$ per locus per generation for the trimmed data (see also Lohse et al~2011). 
 
Simulated pairwise difference data were generated by first simulating the coalescence times of pairs of sequences and then superimposing neutral mutation under the infinite sites model; our R code is included as Supplementary Material. For pairs of sequences from the same species, our simulation algorithm uses the ``shortcut" provided by equations~(8) and~(9) in Wilkinson-Herbots~(2012), exploiting the fact that the distribution of the coalescence time of a pair of sequences sampled from the same species in the IIM model is a mixture of two ``piecewise exponential" distributions (with ``change points" $\tau_1$ and $\tau_0$), eliminating the need to explicitly simulate migration events. 

\section*{Acknowledgments}

I would like to express my sincere thanks to Yong Wang, Jody Hey, 
Konrad Lohse and Nick Barton for the use of their Drosophila data.
Thanks are also due to Paul Northrop and Rex Galbraith for some helpful tips on programming in R, and to Ziheng Yang for some valuable discussions.
This work was supported by the Engineering and Physical Sciences Research Council via an Institutional Sponsorship Award to University College London (grant number EP/K503459/1).

\section*{Supplementary Material}

The R code which fits an IIM model and an isolation model to pairwise difference data from a large number of independent loci is supplied as Supplementary Material. The R code for simulating pairwise difference data under the IIM model is also provided.

\section*{References}

\begin{description}
\item[]Akaike H. 1972. Information theory and an extension of the maximum
likelihood principle. In: Petrov BN, Csaki F, editors. 
Proc. 2nd Int. Symp. Information Theory, Supp. to Problems of Control
and Information Theory. p. 267-281.
\item[]Akaike H. 1974. A new look at the statistical model identification.
IEEE Transactions on Automatic Control AC-19:716-723.
\item[]Andersen LN, Mailund T, Hobolth A. 2014. Efficient computation in the IM model. J Math Biol. 68:1423-51.
\item[]Becquet C, Przeworski M. 2007. A new approach to estimate
parameters of speciation models with application to apes. Genome Res. 17:1505-1519.
\item[]Becquet C, Przeworski M. 2009. Learning about modes of 
speciation by computational approaches. Evolution 63:2547-2562.
\item[]Bird CE, Fernandez-Silva I, Skillings DJ, Toonen RJ. 2012.
Sympatric speciation in the post ``modern synthesis" era of evolutionary 
biology. Evol Biol. 39:158-180.
\item[]Burgess R, Yang Z. 2008. Estimation of hominoid ancestral  
population sizes under Bayesian coalescent models incorporating mutation rate 
variation and sequencing errors. Mol Biol Evol. 25:1979-1994.
\item[]Carstens BC, Stoute HN, Reid NM. 2009. 
An information-theoretical approach to phylogeography. Mol Ecol 18:4270-4282.
\item[]Cox DR. 2006: Principles of Statistical Inference. Cambridge 
University Press.
\item[]Hey J. 2005. On the number of New World founders: a population 
genetic portrait of the peopling of the Americas. PLoS Biol. 3:965-975.
\item[]Hey J. 2010. Isolation with migration models for more than two 
populations. Mol Biol Evol 27:905-920.
\item[]Hey J, Nielsen R. 2004. Multilocus methods for estimating population
sizes, migration rates and divergence time, with applications to the divergence
of Drosophila pseudoobscura and D. persimilis. Genetics 167:747-760.
\item[]Hey J, Nielsen R. 2007. Integration within the Felsenstein equation
for improved Markov Chain Monte Carlo methods in population genetics. 
Proc Natl Acad Sci USA. 104:2785-2790.
\item[]Hobolth A, Andersen LN, Mailund T. 2011. On computing the 
coalescence time density in an isolation-with-migration model with few
samples. Genetics 187:1241-1243.
\item[]Innan H, Watanabe H. 2006. The effect of gene flow on the coalescent
time in the human-chimpanzee ancestral population. Mol Biol Evol. 23:1040-1047.
\item[]Li H, Durbin R. 2011. Inference of human population history
from individual whole-genome sequences. Nature 475(7357):493-496.
\item[]Lohse K, Sharanowski B, Stone GN. 2010. Quantifying the 
Pleistocene history of the oak gall parasitoid {\em cecidostiba fungosa}
using twenty intron loci. Evolution 64:2664-2681.
\item[]Lohse K, Harrison RJ, Barton NH. 2011. A general method for 
calculating likelihoods under the coalescent process. Genetics 189:977-987.
\item[]Maddison WP, Knowles LL. 2006. Inferring phylogeny despite
incomplete lineage sorting. Syst Biol. 55:21-30.  
\item[]Nadachowska K. 2010. Divergence with gene flow - the amphibian
perspective. Herpetological Journal 20:7-15.
\item[]Nielsen R, Wakeley J. 2001. Distinguishing migration from isolation:
a Markov Chain Monte Carlo approach. Genetics 158:885-896.
\item[]Pinho C, Hey J. 2010. Divergence with gene flow: models and
data. Annu Rev Ecol Evol Syst. 41:215-230.
\item[]Powell JR. 1997. Progress and Prospects in Evolutionary Biology:
The Drosophila Model. Oxford University Press.
\item[]R Development Core Team. 2011. R: A language and environment for
  statistical computing. R Foundation for Statistical Computing,
  Vienna, Austria. 
URL http://www.R-project.org/.
\item[]Self SG, Liang KY. 1987. Asymptotic properties of 
Maximum Likelihood Estimators and Likelihood Ratio Tests under non-standard
conditions. J Am Stat Assoc. 82:605-610.
\item[]Smadja CM, Butlin RK. 2011. A framework for comparing
processes of speciation in the presence of gene flow. Mol Ecol. 20:5123-5140.
\item[]Sousa VC, Grelaud A, Hey J. 2011. On the nonidentifiability  
of migration time estimates in isolation with migration models. Molecular
Ecology 20:3956-3962.
\item[]Strasburg JL, Rieseberg LH. 2010. How robust are ``isolation
with migration" analyses to violations of the IM model? A simulation study.
Mol Biol Evol. 27:297-310.
\item[]Strasburg JL, Rieseberg LH. 2011. Interpreting the 
estimated timing of migration events between hybridizing species. 
Molecular Ecology 20:2353-2366.
\item[]Takahata N. 1995. A genetic perspective on the origin and history of
humans. Annu Rev Ecol Syst. 26:343-372.
\item[]Takahata N, Satta Y, Klein J. 1995. Divergence time and population
size in the lineage leading to modern humans. Theor Popul Biol. 48:198-221.
\item[]Takahata N, Satta Y. 1997. Evolution of the primate lineage
leading to modern humans: phylogenetic and demographic inferences from 
DNA sequences. Proc Natl Acad Sci USA. 94:4811-4815.
\item[]Teshima KM, Tajima F. 2002. The effect of migration during the
divergence. Theor Popul Biol. 62:81-95.
\item[]Wang Y, Hey J. 2010. Estimating divergence parameters with
small samples from a large number of loci. Genetics 184:363-379.
\item[]Watterson GA. 1975. On the number of segregating sites in genetical  
models without recombination. Theor Popul Biol. 7:256-276.
\item[]Wilkinson-Herbots HM. 2008. The distribution of the coalescence
time and the number of pairwise nucleotide differences in the "isolation
with migration" model. Theor Popul Biol. 73:277-288.
\item[]Wilkinson-Herbots HM. 2012. The distribution of the coalescence
time and the number of pairwise nucleotide differences in a model of 
population divergence or speciation with an initial period of gene flow.
Theor Popul Biol. 82:92-108.
\item[]Yang Z. 1997. On the estimation of ancestral population sizes of 
modern humans. Genet Res Camb. 69:111-116.
\item[]Yang Z. 2002. Likelihood and Bayes estimation of ancestral 
population sizes in hominoids using data from multiple loci. 
Genetics 162:1811-1823.
\item[]Yang Z. 2010. A likelihood ratio test of speciation with gene flow
using genomic sequence data. Genome Biol Evol. 2:200-211.
\item[]Zhu T, Yang Z. 2012. Maximum Likelihood implementation of an 
isolation-with-migration model with three species for testing speciation
with gene flow. Mol Biol Evol. 29:3131-3142.
\end{description}

\newpage

\end{document}